# A Gossip-Based Optimistic Replication for Efficient Delay-Sensitive Streaming Using an Interactive Middleware Support System

Constandinos X. Mavromoustakis, *Member, IEEE*, and Helen D. Karatza, *Senior Member, IEEE*

*Abstract*—Emerging Mobile-Peer-to-Peer (MP2P) systems involve client applications that dynamically demand resources directly from the global environment. While sharing resources the efficiency is substantially degraded as a result of the scarceness of availability of the requested resources in a multiclient support manner. These resources are often aggravated by many factors like the temporal constraints for availability or node flooding by the requested replicated file chunks. Thus replicated file chunks should be efficiently disseminated in order to enable resource availability on-demand by the mobile users. This work considers a cross layered middleware support system for efficient delay-sensitive streaming by using each device's connectivity and social interactions in a cross layered manner. The collaborative streaming is achieved through the epidemically replicated file chunk policy which uses a transition-based approach of a chained model of an infectious disease with susceptible, infected, recovered and death states. The Gossip-based stateful model enforces the mobile nodes whether to host a file chunk or not or, when no longer a chunk is needed, to purge it. The proposed model is thoroughly evaluated through experimental simulation taking measures for the effective throughput $E_{ff}$ as a function of the packet loss parameter in contrast with the effectiveness of the replication Gossip-based policy.

## I. INTRODUCTION

TODAY'S mobile devices' philosophy of running the 3A (Anything, Anytime, Anywhere) creates a fertile research environment for developing multimedia application like the on-the move video requests by Mobile users. Mobile Peer-to-Peer (MP2P) devices are characterized by bounded resource sharing reliability. Since mobile hosts move freely, network partitioning occurs frequently, and thus data accessibility is relatively lower than that in conventional fixed networks. Due to the asymmetry in wireless communication and the scarceness of wireless resources, there are many open research issues for the techniques which enable high resource availability in mobile peer to peer networks. Particularly in infrastructureless wireless networks when as a mean for disseminating files and enabling the on-demand file sharing transfers, there are many problems that should be faced in order to face the dynamically changing topology and devices' mobility problems that arise.

Dealing with the asymmetry characteristics in wireless devices and the scarceness in wireless resource sharing availability, MP2P networks should form the topology dynamically and in way that the network formation will survive to host the end-to-end resource availability between hosts. Multimedia streaming in such environments is facing reliability problems where usually resources are wasted in the end-to-end connectivity guarantee and network formation. Additionally delay is not optimal due to the instability of the availability of the node involved in the path for the provision of the availability of the "*on the run*" file and requested resources. As a result of all the above there is a severe throughput instability showing often the inadequacy in hosting delay sensitive applications. In a P2P network, some factors aggregate and cause performance reduction. These factors are based on connection quality or node and resource availability or maintenance between peers on the network. Peers are prone to failures, and failures against performance are then aggravated by short connections times or sudden disconnections (with chained unpredictable disconnections due to range and battery failures) and small network formation factor. As a result redundant information is disseminated to all neighboring nodes in order to contribute to network reformation. But this substantially aggravates the overall performance since it generates redundant information that is destined only for a small group of mobile users. Moreover it utilizes the network resources (capacity and channels) causing additional delay which aggravates the overall throughput of the system [1]. In this work a new model for disseminating information is presented where the Epidemic model is used and applied into a MP2P system. Along with the designed model and framework, a throughput model approach is proposed and encapsulated in the epidemic model for primarily enabling the stability in the offered resource reliability and increase the throughput of the system by increasing the throughput per source-destination pair scale with the number of nodes $n$. The proposed model enables the involved nodes in the path to contribute into the diffusion process pathetically and increases the multicasting diffusion throughput response significantly.

## II. RELATED WORK

Nowadays there is an increasing tendency for people to share resources (e.g., files) in a P2P manner, even in a wireless en-vironment, using ad-hoc networking. File sharing is probably the most widespread P2P application. It is estimated that as much as 70% of the network traffic in the Internet can be attributed to the exchange of files, in particular music files [12] (more than one billion downloads of music files can be listed each week [13]). When a mobile node makes an explicit request for a resource the whole network is flooded with a single query, as is the case with many mobile ad-hoc route discovery algorithms [3], [4]. Similar to file discovery by query flooding in P2P networks like Gnutella and unlike the proposed scheme in [5] which enables efficient and consistent access to data the majority of techniques enable redundant message generated communication overhead. Flooding in a wireless network is in fact as efficient as compared to in wired networks because of wireless multicast advantage [6]. Improvements of the basic flooding approach using advertisements and geographic information have also been recently studied [7], [8]. Event dissemination protocols use gossip replicas to carry out multicasts. Different mechanisms to broadcast information in the complete network were proposed in the recent past exploring the advantages in disseminating data in a specific geographic area (Geocast [9]) as well as the caching approaches used, for enabling the requested data content to be available and discoverable [10], [11] at any time such that content can be discovered in a peer-to-peer manner without having network partitioning problems. Additionally if requested data is available the timing end-to-end availability should use the selective or unselective dissemination process to forward the requested packets to destination [1], [2], [10], [11]. However selective and criteria-based dissemination procedure based on mobile nodes' content requirement and the reactive multicast group establishment is still a relatively unexplored area. Proactive multicasting can be disastrous and prohibitive, particularly given the fact that membership varies over time due to dynamically changing topology (frequent and unexpected changes in topology/managing group membership is difficult), the content requirement can be diverse (need for multiple multicast groups), and the asymmetry in nodes' resources which significantly affects the stability (hosting many capacity and energy constraints). In [14] authors identify several interesting effects at the link-layer, notably the highly irregular packet reception contours, the likeliness of asymmetric links, and the complex propagation dynamics of simple protocols. An epidemic algorithm in [15] is proposed based on strictly local interactions for managing replicated databases in a robust way for unpredictable communication failures. An extension of this method is studied in [16] where the "repopulation" process is applied for facing node failures. When a mobile node makes an explicit request for a resource, the whole network is flooded with a query, like mobile ad-hoc route discovery algorithms [17]. Significant improvements of the basic flooding approaches using advertisements and geographic information have also been recently studied [18]. In this work the decentralization and autonomy of control of all the nodes in a cluster which has at least one active transmission (delay sensitive data/multimedia packets requested from a node to another) is proposed and studied using an Epidemic-based replication approach in clustered MP2P devices. Each node in a P2P network can autonomously determine when and to what extent (number of hops) it makes its resources available to other nodes/entities while enabling decentralization. In the proposed model there is no central coordinating authority for the organization of the network (setup aspect) or the use of resources and communication between the peers in the network (sequence aspect). This applies in particular to the fact that no node has central control over the other. In this respect, communication between peers takes place directly. Basically an approach for reliable file sharing based on the advantages of epidemic selective dissemination through temporally-aware cluster-based configuration is introduced. The proposed scheme combines the strengths of both autonomic gossiping and hybrid (nonstatic) Infostation [18] concept and attempts to fill the trade-off between user mobility, reliable file sharing and limited throughput in exchanging delay sensitive data in mobile peer to peer environments. Examination through simulation is performed for the offered reliability by the epidemic collaborative replication scheme and the combined throughput optimization scheme, showing the significant increase in the grade of robustness and the file sharing reliability among mobile peers and a remarkable throughput response.

### III. Gossip-Based Middleware Support System for Collaborative Replication for Maintaining File Sharing Consistency

#### A. Cooperation Scheme Between Clustered Mobile Peer-to-Peer Devices

Infrastructureless wireless networks due to each device asymmetrical limitations have constrained capacities and bounded capabilities in terms of reliable end-to-end connectivity. Thus when a delay sensitive stream is about to be transmitted over different relay forwarding nodes (due to a lack of infrastructure support, each node acts as a router, forwarding data packets for other nodes) there are many constrains like resources' availability and network connectivity issues that arise on a mobility based scenario. The exploitation of the time-variation of the user's channel due to mobility in [19] creates a new concept by splitting the packets of each source node to as many nodes as possible. Therefore, strategies of this type incur additional delay, because packets have to be buffered until the channel becomes sufficiently strong for transmission(s). In this work in order to avoid any redundant transmissions and retransmissions we propose the clustered-based mobility configuration scenario which is set in Figs. 1 and 3. Clusters enable the connectivity between nodes and the local (within a cluster) control of a specified area. The total transfer delay is measured by superpositioning the transfer delays of the relative relay nodal points (Fig. 1). On the contrary with [1], [18] in this work a different mobility scenario is examined where the node controlled area is not specified (unless a cluster can not be formed) -like the Landscape in [1] and the mobility scenario is entirely different having more flexible registering *in* and *out* of a cluster. Cluster network formation works as follows: each cluster is responsible to host newly added nodes and measures (Cluster Head (CH) responsibility) whether these nodes can host new file chunks. If the new node entered the cluster $i$ has available remaining capacity greater than the existing CH, then the Gradual Energy Tree-based (GET) is formed for maintaining connectivity through the node's remaining energy. GET is used to form a tree having as a root, the node with the least remaining energy, and as children the nodes with the higher residual energy. GET configuration is applied only locally, on a 2-hops basis to each node to prevent huge information delivery to nodes which are located far from source node and to maintain 2-hops recursive network topology formation as in the proposed scenario depicted in . No CH has a restriction to directly communicate with any other CH. Additionally the selected CH has as a responsibility to drive the transfers (between nodes) and restrict transfers which may be inadequate in terms of resources (coverage, connectivity, lack of relay nodes, etc). Devices which are moving into different locations where the nodes' density is not adequate to serve the hosted users are not hosted neither as CH nor as a relay node in a specified cluster. Thus in our model we have enabled the *probation slot* parameter which basically evaluates the time the node which has entered the cluster and after $T_s$ probation slot the node can be either CH candidate or a member of a cluster to share resources. Users' request patterns may vary because some users may require transmitting delay sensitive data which in turn require low latency while others may require data being transmitted with low communication overhead. Devices are purely independent in terms of mobility patterns as well as in terms of capacity having spatiotemporal asymmetry. Connectivity can change network state also when a user moves to a different location and data need to be delivered from a source user to another then the relay mechanism can be interrupted and user experiences data losses. Assume that node 1 has a download request from node 15 in different cluster (Fig. 1). The CH that node 1 is set, gets informed about this transfer and supervises the correct delivery and correct connectivity between peer in the cluster A. The same exists for node 15 as well for all nodes which are not set in one of these clusters (via a third cluster/i.e., cluster C). In this way the connectivity can be adequately formed and nodes can transfer data and file chunks by temporarily cache them onto intermediate nodes as explored in the following section. The proposed scheme could also enable scalable broadcast among mobile nodes while reduces the acknowledged delivery of information in topologically changing environments.

### B. Cross Layered Middleware Support System for Efficient Delay-Sensitive Streaming

In order to enable the transmission of delay sensitive streams a Middleware Support System (MSS) is proposed in cooperation with the system proposed in section D through the object replication scheme. The proposed MSS encounters the prioritization of the stream in a stream oriented way as in [2]. Considering that there may be a number of intermediate nodes in-between when transferring delay sensitive streams from a source to a destination, the proposed MSS will enable the minimizations of delays through the relay points $R_1$ to $R_n$ as in Fig. 1.

In the stream oriented approach as introduced in [1] where packets are a part of streams which comprise a file—with file chunk correlations (like in real time multimedia audio and video streams). All packets have a time $\tau$ for reaching destination

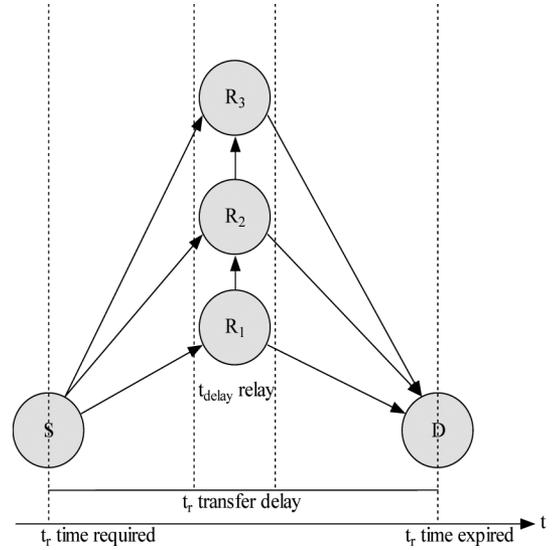

Fig. 1. Relay points on a hop-by-hop basis in Mobile P2P systems.

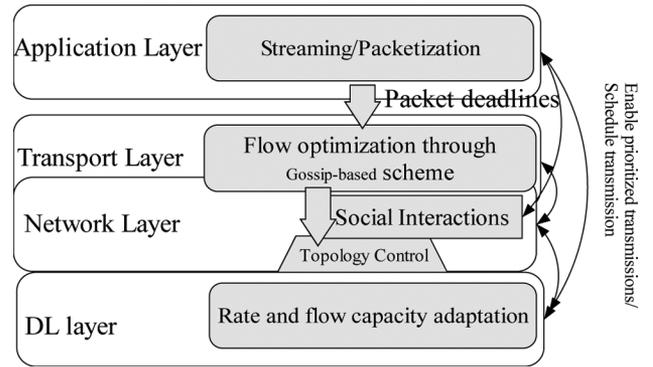

Fig. 2. Cross layered interactive Middleware Support System for efficient delay-sensitive streaming.

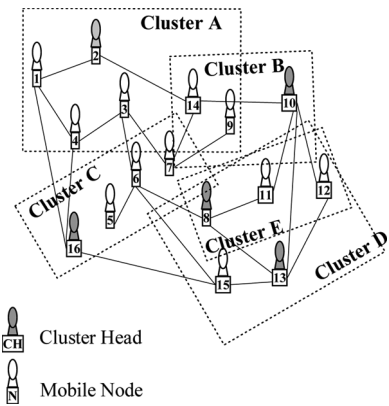

Fig. 3. The initial network formation in a Mobile P2P system according to clustering formation and node self-awareness.

in a free erroneous mode as expressed in [2] in order to arrive correctly at a specified destination. $S_n$ is streaming packet parameter which serves a single application. These packets are marked as prioritized and are considered as delay sensitive. The $S_{t-\tau,j}(S_1, S_2, S_3, \ldots, S_n)$ is called streaming delay bound, where j is the number of the possible intermediate relay nodes, that any of the stream packets $S_n$ might follow, and $S_{t-\tau,j}$ is the upper bound of the required time for correct reception of the stream, at the destination. The prioritized have a bounded time delay $\tau$ to reach any specified destination. The streaming parameter is based on the number of hops from a source to a destination and on the relay region and enclosure graph. The relay region of a certain transmitter-relay node pair (u, w) identifies the set of points in the plane (node locations) for which communicating through the relay node is more reliable than direct communication. Formally,

*Definition 1.1 (Relay Region):* The relay region of a certain transmitter-relay node pair $(u, w)$ is defined as:

$$RR_{u \to w} = \{(x,y) \in \Re^2 : P_{u \to w \to (x,y)} > P_{u \to (x,y)}\} \quad (1.1)$$

where $P_{u \to w \to (x,y)}$ is the probability for a certain node $u$ to transfer the file chunks from a source node $x$ to destination node $y$ via the $w$, based on the connectivity and the social interactions described in the next section. The shape of the relay region depends on the radio signal propagation model, connectivity between nodes and on the value of the distance-connectivity-bounded transfer delay gradient. Thus in an end-to-end path the minimized ping delays between the nodes in the end-to-end path the minimized evaluated delay is according to the:

$$d_p = Min \sum_{i=1}^{n} D_i \quad (1.2)$$

where $D$ is the delay from a node $i$ to node $j$, and $d_p$ is the end-to-end available path.

In Fig. 2 the proposed delay sensitive streaming scheme is mapped onto a cross layer communication stack. In particular, real-time media traffic such as voice and video typically have high data rate requirements and stringent delay constraints, whereas wireless nodes generally have limited or momentarily connectivity. The proposed middleware traversing the Data Link, Network, Transport and Application Layers considers the scarceness of these resources and via the topology control and the gossip-based scheme for maintaining and control the transfer duration it can support delay critical media traffic as the numerical simulation results show. Based on the application, the Application Layer considers the packet classification (delay sensitive or not) and the packets' deadlines in order to enable the Epidemic object replication scheme for reliable file sharing described in section D. Topology control is performed through social interactions expressed in Section III-C, where then the gossip-based mechanism continues to execute during the file sharing process. The topology formation is according to the clustering approach as Fig. 1 shows. The topology is constructed using only local information that preserves the original network connectivity (at least with high probability) using only bidirectional links.

The purpose of using the cross-layer interaction is the defined for each node Distributed State in an infrastructureless-based scheme (MP2P). Unlike traditional infrastructure models where base stations have a global view of network state, the network state in MP2P networks is generally distributed across the nodes. Each node forms its own local view of state, representing a partial view of the overall network state. As such, each node can run the gossip-based algorithm, locally, using its partial view of network state. Based on this the proposed scheme can exploit a cross-layer design to enable each node to perform fine-grained optimizations locally whenever it detects changes in state. The proposed MSS enables the Inherent Layer Dependencies where several interlayer dependencies exist expanded from the data link through the transport to Application merging the chained properties gained by each layer to reflect to the application layer. In essence the proposed approach provides upper protocol layer independency through the MSS.

### C. Considering Social Interactions in the Cross Layered Middleware Architecture for Collaborative Streaming

In this section, we propose a number of social interaction parameters which take place in collaboration with the described cross-layer approach of the previous section. The metrics are community-oriented and are considering the number of created clusters $C_N(t)$ in a specified Relay region of a certain transmitter-and a number of receivers (1, N) under the relay node pair (u, w)—as a modified definition of [26]—as follows (1.3), shown at the bottom of the page, where $W$ is the Community streaming factor and is defined as the number of existing communities in the intercluster communicational links at a given time instant. The $h_N(t)$ is the number of hops in the existing clusters and the $I_{C(N)}(t)$ is the number of interconnected nodes N in the cluster $C_N(t)$. A community is defined as a dense sub-graph where the number of intracommunity edges is larger than the number of intercommunity edges [26]. $W$ can be defined according to the download frequency of the file chunks in the intercommunity as follows:

$$W_N(t) = \frac{DldRate \cdot \#sharingChunks}{Total\#dlds(t) \cdot \#inactiveChunks} \quad (1.4)$$

where in (1.4) the download rate is considered in contrast with the number of chunks being shared in a specified instant time $t$.

*Neighboring Feedback for File Chunk Indices:* Suppose that in the created clusters $C_N(t)$ of specified relay region, there is a $h_N(t) > N(n-1)/2$-as in fully interconnected peer network, for a certain file chunk at a specified time. Then the neighbor $N_j(t)$ at a certain instant of time, informs the k-neighbor receivers for the existence of the file chunk onto this node, according to the following:

$$R_{j \to 1..k} = \left\{ \lim_{N \to K} C_{n(t)} \in W_N(t) : h_N(t) > \frac{N(N-1)}{2} \right\}. \quad (1.5)$$

This means that for a specified amount of time the neighbors collaboratively can provide any node which exists in the Community with a streaming factor $W$, and can be locally informed about any requested file chunk at the specified time $t$. Provided

$$C_N(t) = \frac{2 |h_N(t)|}{|I_{C(N)}(t)| \cdot (|I_{C(N)}(t)| - 1)}, \quad \text{iff } P_{u \to w \to (x,y)} > W_N(t) \quad (1.3)$$

that all the assumptions were made under the epidemic replication scheme described in the next section for enabling reliable file chunks sharing among mobile peers using a stateful metaphor from the epidemics.

### D. Epidemic Optimistic Object Replication Scheme for Reliable File Sharing

Epidemic-like algorithms [20] which follow a nature paradigm could improve the cooperative response of these dynamically changing topologies. By applying simple epidemiological rules to efficiently spread information by only having a local view of the environment, the outcome could in essence a heuristic optimization approach to the reliability problem [10]. Reliable file sharing can be determined by relying on epidemic algorithms, a breed of distributed algorithms that find inspiration in the theory of epidemics. Epidemic (or gossip) algorithms constitute a scalable, lightweight, and robust way of reliable disseminating information to a recipient or group of recipients, by providing guarantees in probabilistic terms. Based on certain characteristics, epidemic algorithms are amenable to the highly dynamic scenarios. In this work a promiscuous caching [1] is used which means that data can be cached "anywhere, anytime". However this enables trade offs in consistency for availability which is faced with cooperative cluster-based connectivity and the caching used in the clusters described earlier. We assume that a common lookup service is followed by all devices in the network cooperating via a shared platform. In a dynamic MP2P network each user might desire to share or download a file or files with other users (peers). Many conditions must be satisfied for reliable communication between mobile peers. On one hand users due to their mobility might draw away from the user (peer) when a file sharing communication takes place. On the other hand a sudden network partitioning or network split could occur because of network's dynamic topology which is continuously changing. Thus a proactive dissemination scheme must be determined in order to prevent the cutoff in file sharing communication. On the contrary with [18] this work assumes an isolated system comprising of a variable number of mobile nodes ([18] assumes a fixed number in a Landscape and pre-known mobility pattern like in VANets) confined in a predefined geographic region. These nodes are mobile, and communicate with each other in a wireless (radio) ad-hoc manner. The limited connectivity coverage that MP2P systems offer, results in significant delay in downloading a message or file (group of packets). In the proposed scenario each node carries some unique data items and the number of nodes which have been requested to deliver file chunks also varies.

A chain model was chosen to determine the file sharing termination criteria as follows: each mobile host $m_k$ has a predetermined capacity $M$. At any time in the network each $m_k$ has a state. There are three different states that $m_k$ can be characterized: the susceptible state $S(t)$ represents the number of hosts in the system which are "susceptible," infected state $I(t)$ represents the number of "infected" hosts, and $R(t)$ represents the "recovered" hosts. A host is in susceptible state $S(t)$ if the device does not share any information with any other host. In turn host/node 1 is in infected state $I(t)$ if a file(s) share occurs. Finally node 1 is in "recovered" state $R(t)$ if any shared file(s) are no longer pending. A chain model of an infectious disease with susceptible, infected, recovered and death states is used shown in Fig. 4.

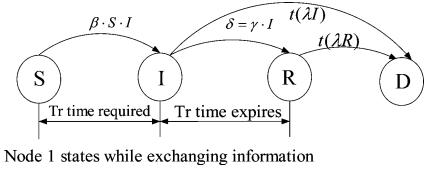

Fig. 4. Chained model of an infectious disease with susceptible, infected, recovered, and death states.

The modeling of diseases implemented for static networks have been studied in the past [20], and this model is used in a similar manner in this mobility scenario. Adopting the framework from an infectious disease model [21], a host is set as "infected" if a file sharing (or a group of data packets) are pending. Suppose there are $k$ hosts in the system, then a host is sharing a resource with $\beta(k-1)$ other hosts per unit time. $S(k-1)$ do not have yet the disease. Therefore, the transition rate from state S to state I becomes

$$Filesharing = (nu\_infected) \cdot (dld\_Rate)$$
$$\cdot (nu\_NOT\_share)$$
$$Filesharing = I\left[\beta(k-1)\right] \cdot \left[\frac{S}{k-1}\right] \quad (1.6)$$

where $\beta$ is the contact rate for $k$ hosts.

Then the downloaded (no longer pending) rate is

$$\delta = \gamma \cdot I \quad (1.7)$$

where $\gamma$ is the download rate and $I$ is the number of infected devices. The file sharing process should be performed in time limit $[0, t)$ and the infection (download) has upper limit time duration otherwise the infection stops and the node is transferred into the recovered state as follows: $I_{TTL} = \tau \cdot \beta \cdot S \cdot I$ (1.3) where $\tau$ is the time limit that a delay sensitive transfer/download can be processed without losses.

In the proposed scheme—on the contrary with [1]—we have introduced the death state which comprises of marking the spread of the deletion of a data item to any nodes which have hosted for time t the data item. Deleted items get the death certificates after a specified timestamp and nodes diffuse this deletion like the updates in any other state (refer to Fig. 4 chain model where an item can not remain after a certain time on a specified node). For this state two thresholds $t1, t2$—most sites delete it after $t1$ and retention sites delete after $t2$.

Assume that a device P has such a certificate for data item x then if by any chance an obsolete update for x reaches P; P will react by simply spreading the death certificate for x again to all nearby nodes in the cluster (informing also the CH). This is to clean up with redundant files the devices' storage units and memories:

$$d_{TTL} = t(\lambda I) + t(\lambda R) = t(\lambda I + \lambda R) = t\lambda(I + R). \quad (2.1)$$

Buffer Purging Enforcement Degree (BPED) as a function of the death rate considered in Fig. 4 (as in the algorithm of Fig. 5), as follows:

$$\text{BPED} = \sum_{t=0}^{t_r} f(\lambda t) \cdot R. \quad (2.2)$$

```
Set communication Peers(A,B);
Call_Index(file); //check via the table lookup search
mechanism
Check_delay(A,B)
While (communication==1){
 If delay criteria meet
    { if (exists(A)==LOCAL_CLUSTER)
       while
((file_chunk==EXISTS)&&(TOTAL_delay()<Tlim(filechunk))){
   if (social_interaction(node A, node B)==K_valid)
    // K_valid ∈ C_N
    {
     Check_CLUSTER(); //initialize the intracluster chunk
sharing
     Start download(source A, dest B);
     If (communication(A,B)!=valid)
         { Check_node();//if A causes the failure or B
           if A causes the failure
           {
            Find_neighbours(A,nu_neighbours[j]);
             For (i=1; i<=nu_neibours; i++){
              Check_capacity(A[j], nu_neighbours);
               nu_neighbours++;
             }
          } if (social_interaction()==K_valid)
              Epidemic_diffusion(A[j]); //selection of
neighbours
              Copy_packets(A[j]);
              Serch_clusterHead(Role_node(A,B),
Intracluster(A,B)
                While (TRUE_IN_CLUSTER(i)){

Epidemic_diffusion(A[Role_node(j)]);//selection of
neighbours
              }//While Role_node
          }//if A causes the failure
```

Fig. 5. Pseudocode for epidemic caching/replication using the time slots and delay mechanisms of the proposed configuration for reliable file sharing between peers.

The total delay of the state-based model can be measured as the subdelays of the measured $T_{tr}$ transferred delay between the states S and I, and the transfer delay between the states I to R and before the transfer time expires as

$$\text{Total D} = \sum_{n(t)=0}^{t_{r(\max)}} T_{tr}(n(t)) \quad (2.3)$$

where the $n$ stands for the hop count measure until reaching the destination and $T_{tr}$ is the mean transferred time at a certain time $t$. The $\beta \cdot S \cdot I$ in (1.3) is called $\pi$ coefficient which indicates the enforcement degree of the diffusion process and $\pi \cdot \tau$ the enforcement at a bounded time delay. $\pi$ has the dimension of $[1/Time]$. Previous examinations of the behavior of small scale systems showed that relatively small populations could be faced with a stochastic model. Thus taking into account that depends on the measures of $S(t)$ and $I(t)$ and the probability of transmitting the information, we can derive $S(t)$ as follows:

$$\frac{dS}{dt} = -\beta \cdot S \cdot I = -\pi$$
$$\frac{dI}{dt} = \beta \cdot S \cdot I = \beta(N-I) \cdot I = \beta NI - \beta I^2. \quad (3.1)$$

By solving the first order differential equation the outcome is

$$I(t) = \frac{N}{1 + e^{-\beta \cdot N \cdot t}(N-1)}. \quad (3.2)$$

According to the definition of spreading ratio (3.2) becomes

$$I'(t) = \frac{I(t)}{N} = \frac{N}{N \cdot (1 + e^{-\beta \cdot N \cdot t}(N-1))}$$
$$I'(t) = \frac{1}{1 + e^{-\beta \cdot N \cdot t}(N-1)} \quad (3.3)$$

Equation (3.3) is referred as the Cumulative Distribution Function (CDF). In the implementations of the scheme the locations will be updated and measured according to the following:

$$L(t) = L(t-1) + S_t \cdot \vec{d}. \quad (4)$$

where $L(t)$ is the new location $L(t-1)$ the previous location at step time $(t-1)$, $S_t$ is the speed of each device and $\vec{d}$ is the directed unit vector [21]. Additionally we estimate the average delay experienced by all the peers in downloading a multipart file as follows:

$$\bar{d}^{(m)} \approx \frac{\tau_0}{m} \log_2 n \quad (5)$$

where $m$ is the number of identical sized chunks that the file is divided, $n$ is the number of peers and $\tau_0$ is the amount of time taken to download the whole file, if downloaded from a single peer.

### E. Enabling High Throughput Using Epidemic Replication

In order to evaluate the modeled approach we have measured the effective throughput $E_{ff}$ as a function of the packet loss parameter and lost packets as shown in (6) at the bottom of the page.

Additionally the measured throughput per source-destination pair scale with the number of nodes n can be evaluated using $Total\_E_{ff} = \sum_1^j (n_i) \cdot (\#of bit \sin filechunk.TransferredSize/Packet$

$$Effective\,Throughput = E_{ff}$$
$$= 1 - (Packet_{loss}) \cdot \left(\frac{Packet\,Transferred\,Size}{Packet\,Transferred\,Time}\right) \cdot \left(\frac{1}{Band\,width}\right). \quad (6)$$

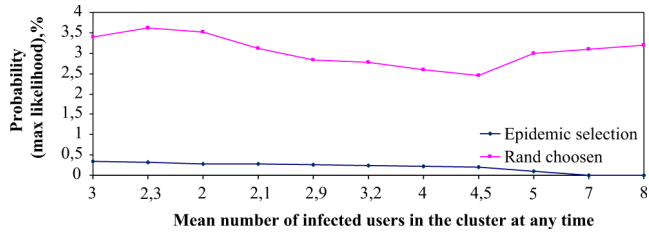

Fig. 6. Mean number of the users that are infected in the cluster.

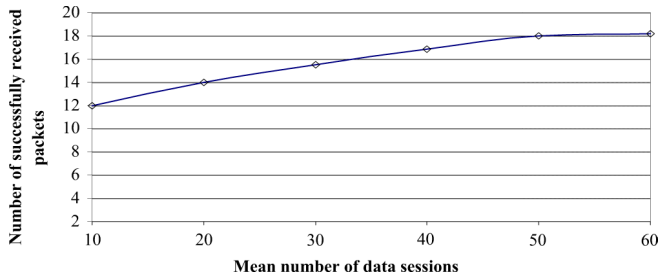

Fig. 7. Mean number of file chunks transfer sessions between clusters, with the number of the successfully received packets(completed downloads).

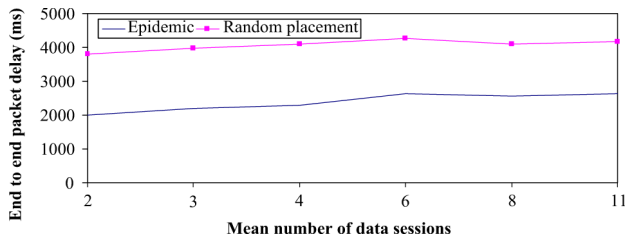

Fig. 8. End to end packet delay (ms) between different clusters, with the mean number of file chunks transfer sessions.

$TransferredTime$), where $n_i$ is the hop metric count (delay measure according to [19], ) from a source to possible destination(s).

## IV. SIMULATION EXPERIMENTS AND DISCUSSION

To demonstrate the methodology discussed in this paper, we performed exhaustive discrete time simulations of the proposed scenario under several different conditions. In the implementation-simulation of this work we used our own libraries implemented in C/Objective C programming language. We assume a system consisting of several mobile nodes, e.g., mobile users equipped with notebooks or PDAs and wireless network interfaces. All mobile nodes collaborate via a shared application that uses a distributed lookup service. Radio coverage is small compared to the area covered by all nodes, so that most nodes cannot contact each other directly. Additionally, we assume IEEE 802.11x as the underlying radio technology. However, it is necessary to point out that communication and epidemic-like dissemination could be employed on any radio technology that enables broadcast transmissions inside a node's radio coverage. In the implementation of the proposed scenario a combination of the Zone Routing Protocol (ZRP) [22] is used with the Cluster-based Routing Protocol (CRP) marking each cluster as a zone. The ZRP is considered advantageous because allows to a certain node to accurately know the neighbors of any mobile terminal within a specified cluster and in a determined number of hops.

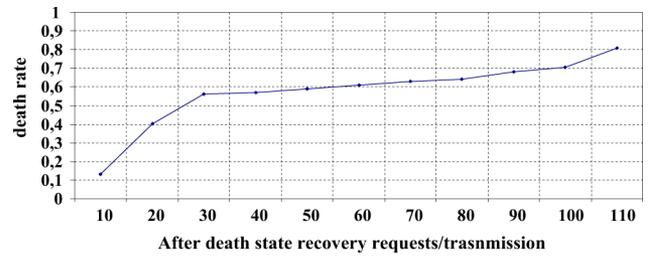

Fig. 9. Death rate with the mean number of file chunks recovery requests.

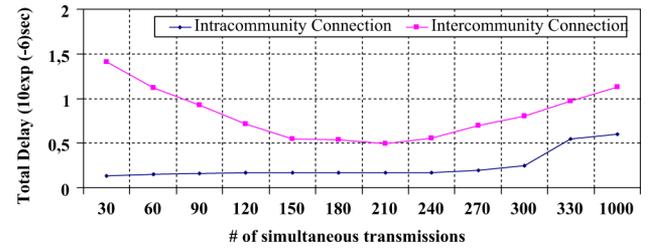

Fig. 10. Total Transfer Delay with the mean number of file chunks simultaneous transmissions (intracommunity and intercommunity).

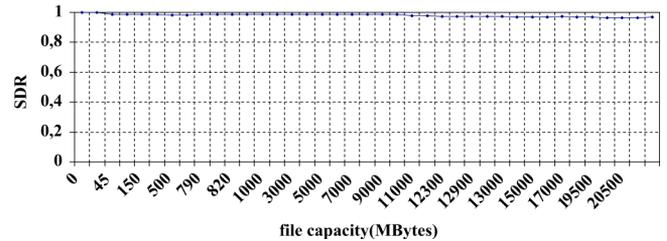

Fig. 11. Successful Delivery Rate (SDR) with the mean capacity of the files which are in active downloads (with their multiple chunks existing onto different nodes).

In Fig. 6 the mean number of the users that are infected in the cluster is shown, for two different selection strategies (epidemic and randomly chosen). Fig. 6 depicts that the proposed epidemic state model can bias in a heuristic way the "shared" users. As seen the mean number of infected users for which the epidemic algorithm is used, is negligible compared with the number of randomly chosen. Fig. 7 shows the mean number of file chunks transfer sessions between clusters, with the number of the successfully received packets of completed downloads. This figure shows that even in the present of high mean number of file chunks transfer sessions between clusters, the packets that are successfully received are kept at high levels.

In Fig. 8, the end to end packet delay (ms) between different clusters with the mean number of file chunks transfer sessions is presented. It is obvious that the epidemic collaboration enables significantly less end-to-end packet delay, even if the number of file chunks transfer sessions increases. The death rate with the mean number of file chunks recovery requests is presented in Fig. 9. If the death rate will not be present this can be in a sense considered as a delay factor [1] (searching delay onto nodes) and as unneeded/redundant information hosted onto nodes. Fig. 8 shows the total transfer delay with the mean number of file chunks simultaneous transmissions. Fig. 8's outcome is considered very interesting extracting that even in the presence of multiple simultaneous transmissions, the total transfer delay is kept relatively at low levels.

The Successful Delivery Rate (SDR) with the mean capacity of the files which are actively downloading their multiple

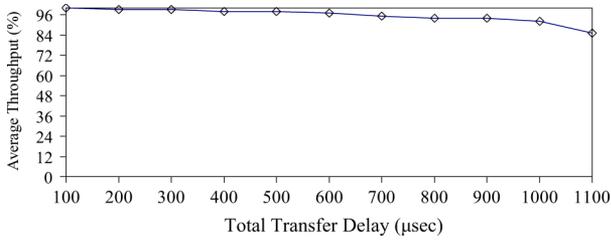

Fig. 12. Throughput evaluation with the Mean Total Transfer Delay (mean of all transfers for the total delay).

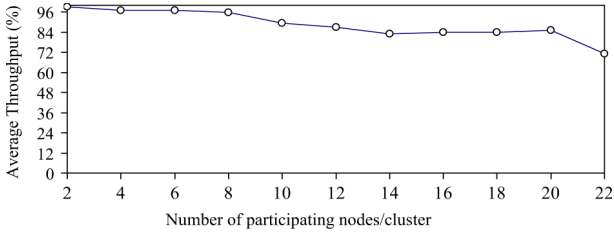

Fig. 13. Throughput evaluation with the mean number of nodes that are participating in the cluster in the transferring (or propagating file chunks).

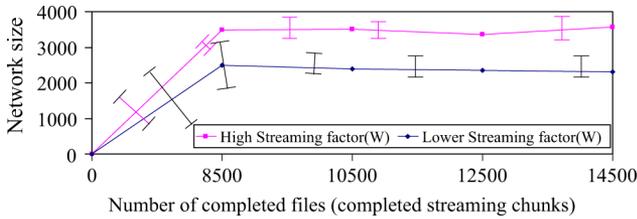

Fig. 14. Network size with the number of completed files.

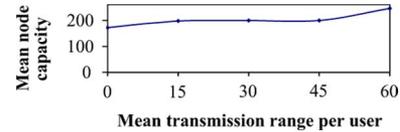

Fig. 15. Mean node density with the mean transmission range.

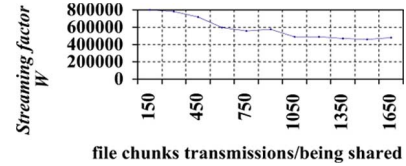

Fig. 16. Streaming factor with the number of file chunks transmissions/being shared.

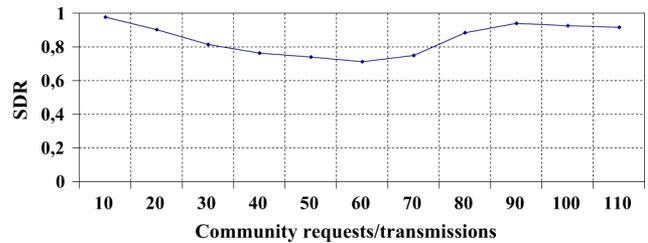

Fig. 17. SDR with the number of the requests in a formed Community.

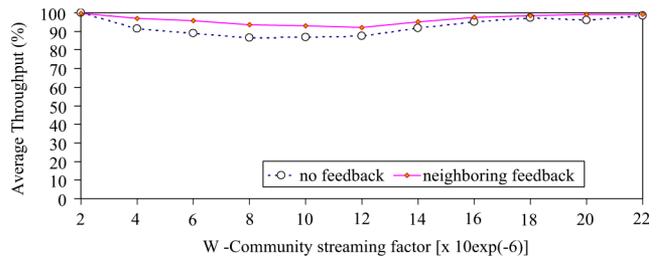

Fig. 18. Average throughput with the Community Streaming factor-$W$.

chunks that exist onto different nodes, is presented in Fig. 11. It is shown that throughout simulation time the SDR drops slightly using the epidemic placement, even when the injected packets reserve great capacities. After consecutive simulation experiments the extracted values did not dropped below 98.1% for successful delivery of the packets. Figs. 12 and 13 show the Throughput evaluations under the Mean Total Transfer Delay (mean of all transfers for the total delay) and the Mean number of nodes that are participating in the cluster in the transferring process (or propagation of file chunks) respectively. Both figures along with Figs. 10 show the robustness of the throughput response under these measures. When the Total Transfer Delay is increasing, the scheme is shown to be robust in the overall throughput offered during the simulation.

Fig. 14 shows the network size with the number of completed file chunks for two levels of streaming (high and low) factors. Since the network size aggravates the response of the network in regards with the streaming factor, Fig. 14 shows that the proposed scheme enables the higher possible network dimensions utilizing the high streaming factor values in contrast with the network nodal dimensions. It is therefore found to be competitively high in terms of response compared with the low streaming factor for the number of completed file chunks. Fig. 15 shows the mean node density with the mean transmission range and Fig. 16 shows the streaming factor with the number of file chunks transmissions/being shared.

In Fig. 17, the SDR with the number of the requests in a formed Community is presented. The SDR is found to be significantly high particularly when high percentage demands exist. The Community streaming factor $W$ is defined as the number of existing communities in the intercluster communicational links at a given time instant. According to Fig. 17 this factor is found to be crucial for the appropriate successful file chunk streaming. Additionally, in Fig. 18, the $h_N(t)$ which is the number of hops in the existing clusters and the $I_{C(N)}(t)$ which is the number of interconnected nodes N in the cluster $C(N)(t)$, seem to positively affect the average throughput which is considered to be significantly and constantly high, particularly in the present of neighboring feedback.

## V. CONCLUSIONS

In this paper, we have proposed an approach for reliable file chunks sharing among mobile peers using a stateful metaphor from epidemics hosted in a MSS. The hosted scheme in the MSS has reflective characteristics in considering the social interactions between mobile peers in cross layered middleware architecture for collaborative streaming. The cross layered interactive

MSS enables through the epidemic scheme efficient delay-sensitive streaming by creating replicas of the file chunks in order to enable P2P reliable file sharing between mobile users. In order to clean up the nodes/users hosting no longer needed and redundant files, we have introduced the Buffer Purging Enforcement Degree (BPED) as a function of the death rate. The proposed scheme seems to behave excellently, allowing high SDR for completed files. A comparison with other schemes for file chunks or generic object's placement is one of our upcoming research targets. The findings of comparing the placement strategy will enable future researches and implementations to use this method in large lookup databases, in which a common multi-client application is used.

Current and future research directions include the modeling of the mobility pattern of the peers by using approaches like the fractional Brownian motion. Additionally the extended scheme will be able to allocate and balance resources among different traffic classes to accomplish the best usage of network resources while maintaining the topology and the wireless connectivity of the users.

## REFERENCES

[1] C. X. Mavromoustakis and H. D. Karatza, "Under storage constraints of epidemic backup node selection using HyMIS architecture for data replication in mobile peer to peer networks," *J. Syst. Softw.*, vol. 81, no. 1, pp. 100–112, Jan. 2008.

[2] C. X. Mavromoustakis and H. D. Karatza, "End-to-end layered asynchronous scheduling scheme for energy aware QoS provision in asymmetrical wireless devices," in *IEEE Int. EDOC Conf. Enterprise Computing Conference 2008, AQuSerM: Advances in Quality of Service Management Workshop*, Munchen, Germany, Sep. 18, 2008, pp. 96–104.

[3] X. Hong, K. Xu, and M. Gerla, "Scalable routing protocols for mobile ad hoc networks," *IEEE Network Mag.*, vol. 16, no. 4, pp. 11–21, Jul./Aug. 2002.

[4] E. Royer and C. Toh, "A review of current routing protocols for ad-hoc mobile wireless networks," *IEEE Pers. Commun.*, vol. 6, no. 2, pp. 46–55, Apr. 1999.

[5] M. K. Aguilera, A. Merchant, M. Shah, A. C. Veitch, and C. T. Karamanolis, "Sinfonia: A new paradigm for building scalable distributed systems," in *21st ACM Symp. Operating Systems Principles (SOSP)*, Stevenson, WA, Oct. 14, 2007, pp. 159–174.

[6] J. E. Wieselthier, G. D. Nguyen, and A. Ephremides, "Energy-efficient broadcast and multicast trees in wireless networks," *Mobile Netw. Applicat. Arch.*, vol. 7, no. 6, pp. 481–492, Dec. 2002.

[7] J. Tchakarov and N. Vaidya, "Efficient content location in wireless ad hoc networks," in *IEEE Int. Conf. Mobile Data Management (MDM)*, 2004, pp. 74–85.

[8] C. X. Mavromoustakis and H. D. Karatza, "Dispersed information diffusion with level and schema-based coordination in mobile peer to peer networks," *Cluster Comput. (Comput. Commun. Netw.)*, vol. 10, no. 1, pp. 33–45, Mar. 2007.

[9] Y. B. Ko and N. H. Vaidya, "Flooding-based geocasting protocols for mobile ad hoc networks," *Mobile Netw. Applicat.*, vol. 7, no. 6, pp. 471–480, 2002.

[10] T. Hara, "Effective replica allocation in ad hoc networks for improving data accessibility," in *Proc. IEEE INFOCOM*, 2001, pp. 1568–1576.

[11] N. Davies, K. Cheverst, K. Mitchell, and A. Friday, "Caches in the air: Disseminating tourist information in the guide system," in *Proc. Second IEEE Workshop on Mobile Computing Systems and Applications*, Feb. 25–26, 1999, pp. 11–19.

[12] M. Stump, Peer-to-peer Tracking Can Save Cash: Ellacoya Nov. 15, 2008 [Online]. Available: http://www.ellacoya.com/news/pdf/10_07_02_mcn.pdf

[13] F. Oberholzer and K. Strumpf, The Effect of File Sharing on Record Sales—An Empirical Analysis Nov. 15, 2004 [Online]. Available: http://www.unc.edu/~cigar/papers/FileSharing_March2004.pdf.

[14] D. Ganesan, B. Krishnamachari, A. Woo, D. Culler, D. Estrin, and S. Wicker, Complex Behavior at Scale: An Experimental Study of Low-Power Wireless Sensor Networks Univ. Caifornia, Los Angeles, 2002, UCLA/CSD-TR 02-0013.

[15] A. Demers, D. Greene, C. Hauser, W. Irish, and J. Larson, "Epidemic algorithms for replicated database maintenance," in *Proc. 6th Annu. ACM Symp. Principles of Distributed Computing*, 1987, pp. 1–12.

[16] J. Kulik, W. R. Heinzelman, and H. Balakrishnan, "Negotiation-based protocols for disseminating information in wireless sensor networks," *Wireless Netw.*, vol. 8, no. 2–3, pp. 169–185, 2002.

[17] X. Hong, K. Xu, and M. Gerla, "Scalable routing protocols for mobile ad hoc networks," *IEEE Network Mag.*, vol. 16, no. 4, pp. 11–21, Jul./Aug. 2002.

[18] C. Mavromoustakis and H. Karatza, "Reliable file sharing scheme for mobile Peer-to-Peer users using epidemic selective caching," in *Proceedings of IEEE Int. Conf. Pervasive Services (ICPS)*, Santorini, Greece, Jul. 2005, pp. 169–177.

[19] M. Grossglauser and D. Tse, "Mobility increases the capacity of ad hoc wireless networks," *Proc. IEEE Infocom*, pp. 312–319, 2001.

[20] C. Bettstetter, "On the minimum node degree and connectivity of a wireless multihop network," in *Proc. 3rd ACM MobiHoc*, Lausanne, Switzerland, 2002, pp. 80–91.

[21] F. Brauer and C. Ch/avez, *Mathematical Models in Population Biology and Epidemiology*. New York: Springer-Verlag, 2001.

[22] Z. Haas and M. Pearlman, "The performance of query control schemes for the zone routing protocol," *ACM/IEEE Trans. Netw.*, vol. 9, no. 4, pp. 427–438, 2001.

[23] C. Plattner, G. Alonso, and M. Özsu, "Dbfarm: A scalable cluster for multiple databases," in *Middleware, 2006.Book Series, Lecture Notes in Computer Science*. Berline/Heidelberg, Germany: Springer, 2006, vol. 4290/2006, pp. 180–200.

[24] F. Picconi and L. Massoulie, "Is there a future for mesh-based live video streaming?," *Peer-to-Peer Comput.*, pp. 289–298, 2008.

[25] P. Ulanovs and E. Petersons, "Modeling methods of self-similar traffic for network performance evaluation," *Scientific Proc. RTU. Series 7. Telecommunications and Electronics*, 2002.

[26] M. Fiore and J. Harri, "The networking shape of vehicular mobility," *Proc. ACM MobiHoc*, pp. 261–272, 2008.

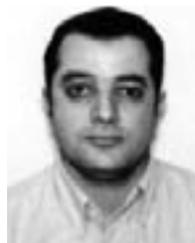

**Constandinos X. Mavromoustak** (M'05) received the five-year dipl.Eng. degree in electronic and computer engineering from Technical University of Crete, Greece, the M.Sc. degree in telecommunications from University College of London, U.K., and the Ph.D. degree from the Department of Informatics, Aristotle University of Thessaloniki, Greece.

He is currently an Assistant Professor, Department of Computer Science, University of Nicosia, Cyprus. He has worked as a Scientific Researcher and as a Research Scientist at the Computer Architecture and Communications Lab and at the Parallel and Distributed Systems Group at the Department of Informatics. His research interests are in the areas of spatio–temporal scheduling in distributed systems, distributed diffusion algorithms, and multidimensional mobility modeling.

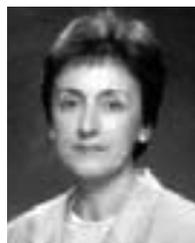

**Helen D. Karatza** (M'95–SM'06) is a Professor in the Department of Informatics at the Aristotle University of Thessaloniki, Greece. Her research interests include computer systems modelling and simulation, performance evaluation of parallel and distributed systems, resource allocation and scheduling, cluster computing, grid computing, energy-efficient scheduling in large scale distributed systems and the grid.